\documentclass[aps,prd,reprint,floatfix,superscriptaddress,lengthcheck,sort&compress,letterpaper,nofootinbib]{revtex4-1}

\usepackage{amssymb}
\usepackage[intlimits]{amsmath}
\usepackage{amstext} 
\usepackage{amsfonts}
\usepackage{dsfont}
\usepackage{float}
\usepackage{slashed}
\usepackage{subfigure}
\usepackage{natbib}
\usepackage{multirow}
\usepackage{verbatim}
\usepackage{nicefrac}
\usepackage{dcolumn}
\usepackage{units}
\usepackage{bm}
\usepackage{wasysym}
\usepackage{array}
\usepackage{graphicx}
\usepackage[usenames]{color}
\usepackage[colorlinks=true,linkcolor=blue,citecolor=blue,urlcolor=blue]{hyperref}

\newcommand{\conjg}[1]{\ensuremath{\hspace{1pt}\overline{\hspace{-1pt}#1\hspace{-1pt}}}\hspace{1pt}}

\def\mS{\ensuremath{\mathcal{S}}}
\def\mD{\ensuremath{\mathcal{D}}}
\def\mT{\ensuremath{\mathcal{T}}}
\def\mC{\ensuremath{\mathcal{C}}}
\def\mF{\ensuremath{\mathcal{F}}}

\begin{document}

\title{$X(3872)$ as a four-quark state in a Dyson-Schwinger/Bethe-Salpeter approach}

\author{Paul C. Wallbott}
\email[e-mail: ]{paul.wallbott@physik.uni-giessen.de}
\affiliation{Institut f\"ur Theoretische Physik, Justus-Liebig Universit\"at Gie{\ss}en, 35392 Gie{\ss}en, Germany}

\author{Gernot Eichmann}
\email[e-mail: ]{gernot.eichmann@tecnico.ulisboa.pt}
\affiliation{CFTP, Instituto Superior T\'ecnico, Universidade de Lisboa, 1049-001 Lisboa, Portugal}
\author{Christian S. Fischer}
\email[e-mail: ]{christian.fischer@physik.uni-giessen.de}
\affiliation{Institut f\"ur Theoretische Physik, Justus-Liebig Universit\"at Gie{\ss}en, 35392 Gie{\ss}en, Germany}
  \begin{abstract}
  We generalise the framework of Dyson-Schwinger and Bethe-Salpeter equations for four-quark states
  to accommodate the case of unequal quark masses. As a first application, we consider the
  quantum numbers $I(J^{PC})=0(1^{++})$ of the $X(3872)$ and study the four-quark states with quark contents
  $cq\bar{q}\bar{c}$ and $cs\bar{s}\bar{c}$. Their Bethe-Salpeter amplitudes are represented
  by a basis of heavy-light meson-meson, hadro-charmonium and diquark-antidiquark operators, which
  allows for a dynamical distinction between different internal configurations.
  In both cases we find the heavy-light meson-meson component to be dominant.
  For the putative $X(3872)$ we obtain a mass of $3916(74)$~MeV;
  the corresponding $cs\bar{s}\bar{c}$ state is predicted at $4068(61)$~MeV.
  \end{abstract}

\maketitle

\section{Introduction}

With the spectacular success of Belle, BaBar, BES III and the LHC experiments and their
discovery of an ever increasing and largely unexplained number of potentially exotic states,
hadron spectroscopy in the heavy-quark region has become a fascinating topic in the past
years, see e.g.
\cite{Esposito:2016noz,Lebed:2016hpi,Chen:2016qju,Ali:2017jda,Guo:2017jvc,Olsen:2017bmm} for
recent reviews.
This started out in 2003, when the Belle collaboration found an unexpected and
surprisingly narrow state in the J/$\psi \pi $\textsuperscript{+}$\pi $\textsuperscript{$-$}
invariant mass spectrum called the X(3872) \cite{Choi:2003ue}. Since then this state has also been seen
by other experiments \cite{Acosta:2003zx,Abazov:2004kp,Aubert:2004ns,Aaij:2011sn}
and its original quantum number assignment of $J^{PC}=1^{++}$ was later confirmed
by the LHCb collaboration \cite{Aaij:2013zoa}.
It turned out that the X(3872) is hard to reconcile with a conventional charmonium meson.
Its position very close to the $D\overline{D}$ threshold is remarkable as well as the
fact that it decays into J/${\psi}\,\pi$\textsuperscript{+}$\,\pi $\textsuperscript{{}--}
and J/${\psi}\,\pi $\textsuperscript{+}$\,\pi $\textsuperscript{{}--}$\,\pi$\textsuperscript{0}
with similar rates. The study of this object is a challenge for both theoretical and
experimental groups; a detailed line shape analysis will be possible in the future PANDA
experiment at FAIR~\cite{PANDA:2018zjt}.

From a theoretical perspective there are many open questions concerning the internal structure
of exotic states. In this work we focus on four-quark states, i.e., states consisting of
two quarks and two anti-quarks in an overall color-singlet configuration. Moreover we specialize
on configurations with a heavy $c\bar{c}$ pair and a strange or light $q\bar{q}$ pair of valence quarks.
These four quarks may (or may not) arrange themselves in sub-clusters. When the mass of the
four-quark state is close to an open charm threshold it is plausible to assume an internal structure
of a {\it meson molecule} \cite{Guo:2017jvc}. The four quarks then arrange themselves into
pairs of $D^{(*)}/\conjg{D}^{(*)}$ mesons interacting with each other by short- and/or long-range
forces. Naturally, such a picture makes most sense if the width of the constituents is smaller
than the width of the four-quark state. The above-mentioned X(3872) has been considered as
a prime candidate for such a state. Another possibility is the internal structure of a
{\it hadro-quarkonium}~\cite{Voloshin:2007dx}, where the heavy quark and antiquark group together in a
core surrounded by the lighter $q\bar{q}$ pair. This picture is motivated by the observation
that several potentially exotic hadrons were only discovered in final states of a specific
charmonium state with light hadrons. It then seems natural to assume that the decay products are already
pre-formed inside the four-quark state. Finally, four-quark states have been described
as bound objects clustered in \textit{diquark-antidiquark} (\mbox{$dq$-$\conjg{dq}$}) components\footnote{Sometimes in the literature
the term 'tetraquark' is reserved for these configurations only, sometimes it is used for
any four-quark state regardless of its internal structure. In this work we adopt the latter
terminology.} \cite{Esposito:2016noz}.
In principle, the three different possibilities of internal clustering are not mutually exclusive.
It may be that some of the experimentally observed states fall into one of the three categories,
whereas others fall in another. It is therefore vital to develop theoretical approaches that can
deal with all of the different possibilities.

Most effective field theory and model approaches to meson molecules, hadro-quarkonia
and \mbox{$dq$-$\conjg{dq}$} tetraquarks  already assume a certain internal structure from the start.
This is different for lattice calculations, which work directly with the
underlying QCD Lagrangian; see e.g.
\cite{Prelovsek:2010kg,Abdel-Rehim:2014zwa,Lee:2014uta,Prelovsek:2014swa,
Padmanath:2015era,Francis:2016hui,Bicudo:2017szl,Francis:2018jyb,Leskovec:2019ioa} and references
therein. Lattice calculations of four-quark states are extremely expensive and thus it seems fair
to say that at least in the charm-quark energy range they are still performed at an exploratory
level at small volumes, coarse lattices and using light quarks with masses larger than the
physical point. Nevertheless, a number
of interesting observations have been made~\cite{Prelovsek:2014swa,Padmanath:2015era}:
(i) in the isospin $I=0$ channel corresponding to the experimentally observed X(3872) a state has
been found, whereas a corresponding state in the $I=1$ channel is absent; (ii) diquark interpolating
operators have been found to play a negligible role, whereas the presence of $c\bar{c}$ and
$D\conjg{D}^*$ operators were crucial.

In this work we present another approach that is able to deal with different internal configurations
in one framework. Ref.~\cite{Eichmann:2015cra} employed the functional approach of Dyson-Schwinger equations (DSEs)
and a four-body Bethe-Salpeter equation (BSE) to describe the lowest
scalar meson octet and successfully reproduced the mass hierarchy of the $f_0(500)$, the $\kappa$ and the $f_0/a_0(980)$.
To this end, a special role of internal meson-meson configurations in the pseudoscalar
meson channels has been identified: The strong binding in these channels due to dynamical
chiral symmetry breaking induces a drastic reduction of the mass of the four-body states from the
natural scale of 1300--1500 MeV (four valence quarks) down to a mass of roughly 400--500 MeV for
the $f_0(500)$. The dominant role of meson-meson configurations also entails that \mbox{$dq$-$\conjg{dq}$}
clusters were found to have a negligible impact.

In the present work we generalise the framework of Ref.~\cite{Eichmann:2015cra} in two respects. First,
we consider quarks with unequal masses, thus accommodating the cases of
$cq\bar{q}\bar{c}$ and $cs\bar{s}\bar{c}$ quark flavours. Secondly, we consider the quantum
numbers $1^{++}$ of the X(3872) channel. We study this state in a spin-flavour basis which
includes all three internal configurations: pairs of heavy-light mesons, hadro-charmonium and \mbox{$dq$-$\conjg{dq}$},
letting the dynamics decide which of these structures is favoured.

The paper is organised as follows. Section \ref{secII} deals with the technical setup of the
framework. We briefly summarise the derivation of the four-body equation, discuss the truncation
of the two-body interactions and the construction of the basis for the Bethe-Salpeter amplitude.
Some technical details are relegated to an appendix.
In Sec.~\ref{secIII} we then discuss our results before we conclude in Sec.~\ref{sec:conclusion}.

\section{Four-quark states in the DSE/BSE approach}\label{secII}

\subsection{Four-body Bethe-Salpeter equation}

\begin{figure}[t]
  \includegraphics[width=1\columnwidth]{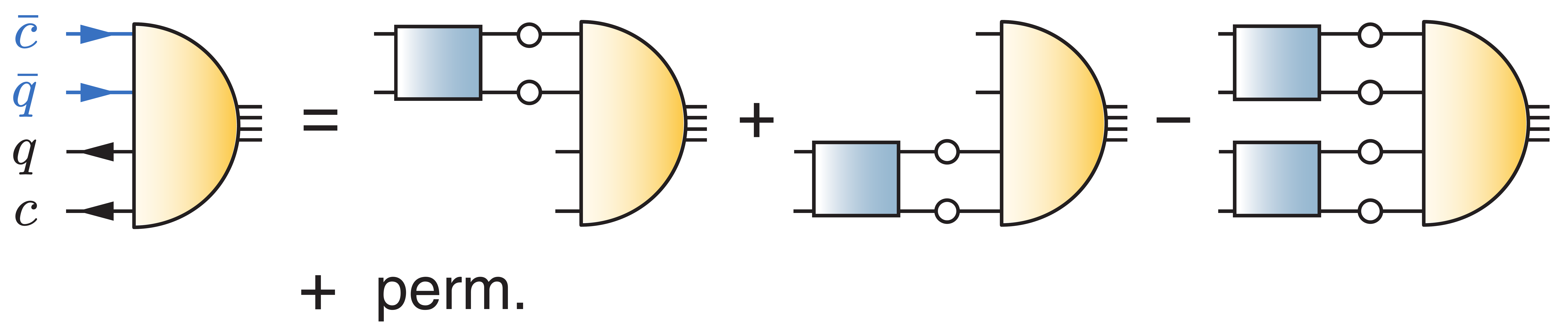}
  \caption{
  Four-quark BSE for a $cq\bar{q}\bar{c}$ system in the $(12)(34)$ configuration; the remaining $(13)(24)$ and $(14)(23)$ permutations
  are not shown. The half-circles and boxes represent the tetraquark amplitude and Bethe-Salpeter kernel, respectively.
  \label{fig_q4_bse}}
  \end{figure}

  Four-quark states in QCD must appear as poles in the
  $qq\bar{q}\bar{q}$ scattering matrix $T$, which is an eight-point correlation function
  and satisfies the scattering equation
  \begin{equation}\label{scattering-eq}
     T = K + K G_0\,T\,.
  \end{equation}
  Here, $K$ is the four-quark interaction kernel and $G_0$ the product of four
  dressed (anti-)quark propagators. In this compact notation, each multiplication
  represents an integration over all loop momenta. The poles in the scattering matrix appear
  for real or complex values of the total momentum transfer $P^2$.
  At a given pole, the residue of the scattering equation is the homogeneous BSE for the
  four-quark BS amplitude $\Gamma$ shown in Fig.~\ref{fig_q4_bse}:
  \begin{equation}\label{eq_bse}
     \Gamma = K G_0 \,\Gamma\,.
  \end{equation}

  In general, Eq.~\eqref{eq_bse} is an eigenvalue equation for $KG_0$
  whose eigenvalues $\lambda_i(P^2)$ depend on $P^2 \in \mathds{C}$.
  If an eigenvalue satisfies $\lambda_i(P_i^2) = 1$, this corresponds to a pole in the scattering matrix
  with $P_i^2 = -M_i^2$.
  If the mass $M_i$ is real and lies below a given meson-meson threshold, it describes a bound state;
  for a resonance the condition can only be satisfied in the complex plane of a higher Riemann sheet.
  Either way, in principle the homogeneous BSE is suitable to detect both bound states and resonances,
  although the calculation of $\lambda_i(P^2)$ above the lowest threshold requires contour deformations
  and direct access to the second sheet depends on knowledge of the full scattering amplitude~\cite{Eichmann:2019dts}.

  The exact kernel $K$ in Eq.~\eqref{scattering-eq} is the sum of two-, three- and four-body irreducible
  interactions. In the following we neglect three- and four-body forces, so that
  the resulting kernel is the sum of two-body interactions:
  \begin{equation}
     KG_0 = \sum_{aa'} K_{aa'}\,,
  \end{equation}
  where $a$, $a'$ denote $qq$, $\bar{q}\bar{q}$ or $q\bar{q}$ pairs and $aa'$ is one of the three
  combinations $(12)(34)$, $(13)(24)$ or $(14)(23)$. $K_{aa'}$ then describes the component of the
  four-body kernel where all interactions are switched off except those within the pairs $a$ and $a'$.
  The resulting equation
  is shown in Fig.~\ref{fig_q4_bse} and can be rewritten as a Faddeev-Yakubovski equation~\cite{Yakubovsky:1966ue}.
  The particular form of the kernel
  \begin{equation}
    K_{aa'} = K_a + K_{a'} - K_a\,K_{a'}
  \end{equation}
  avoids overcounting and ensures the separability
  of the four-body correlation function obtained from one channel $aa'$ only and thus the absence
  of residual color forces between widely separated clusters~\cite{Huang1975,Khvedelidze:1991qb,Heupel:2012ua}.

  We employ a rainbow-ladder kernel for the $q\bar{q}$ and $qq$ interaction, which amounts to an iterated
  dressed gluon exchange and has been reviewed recently~\cite{Eichmann:2016yit} together with more advanced schemes;
  see also~\cite{Sanchis-Alepuz:2017jjd,Eichmann:2019tjk} for details.
  The construction satisfies chiral constraints such as the Gell-Mann-Oakes-Renner relation
  and ensures the (pseudo-)\,Goldstone-boson nature of the pion
  and it has been extensively applied to meson and baryon phenomenology.
  To this end one defines an effective interaction $\alpha(k^2)$
  which incorporates dressing effects of the gluon propagator and quark-gluon vertex~\cite{Maris:1999nt};
  once specified, all further elements of the calculation follow.
  We solve the quark DSE for a range of quark masses and their resulting dressed propagators
  enter in the kernel $KG_0$. The input current-quark masses for up/down, strange and charm
  are listed in Table~\ref{tab:rl}.

\subsection{Four-quark amplitude}

  The BS amplitude of a four-quark state
  with quantum numbers $J^{PC} = 1^{++}$ is the direct product of Dirac, color and flavor parts:
  \begin{equation}
     \Gamma^\mu(p,q,k,P)= \Gamma^\mu_\text{D}(p,q,k,P) \otimes \Gamma_\text{C} \otimes \Gamma_\text{F}\,.
  \end{equation}
  From $\mathbf{3}\otimes \mathbf{3} \otimes \mathbf{\bar{3}} \otimes \mathbf{\bar{3}} =
       ( \mathbf{\bar{3}} \oplus \mathbf{6}) \otimes (\mathbf{3}\oplus\mathbf{\bar{6}}) = \mathbf{1} \oplus \mathbf{1} \oplus ... $, the \\
  color part of the amplitude
  consists of two independent color singlet tensors, which can be picked from
  the \mbox{$dq$-$\conjg{dq}$} ($\mathbf{\bar{3}} \otimes \mathbf{3}$, $\mathbf{6}\otimes\mathbf{\bar{6}}$)
  or either of the meson-meson configurations ($\mathbf{1} \otimes \mathbf{1}$, $\mathbf{8} \otimes \mathbf{8}$),
  for example:
  \begin{equation}\label{color-11}
  \begin{split}
     (\mC_{11})_{ABCD} &=  \frac{1}{3}\,\delta_{AC}\,\delta_{BD} \\
     (\mC_{11}')_{ABCD} &=  \frac{1}{3}\,\delta_{AD}\,\delta_{BC}\,.
  \end{split}
  \end{equation}
  The two tensors in the \mbox{$dq$-$\conjg{dq}$} decomposition are linear combinations of these,
  \begin{equation}\label{color-33}
     \mC_{\bar{3}3} = -\frac{\sqrt{3}}{2}\,(\mC_{11}-\mC_{11}')\,, \quad
     \mC_{6\bar{6}} = \sqrt{\frac{3}{8}}\,(\mC_{11}+\mC_{11}')\,,
  \end{equation}
  as well as the remaining octet-octet tensors:
  \begin{equation}
     \mC_{88} = \frac{\mC_{11} - 3\,\mC_{11}'}{2\sqrt{2}}\,, \quad
     \mC_{88}' = \frac{\mC_{11}' - 3\,\mC_{11}}{2\sqrt{2}}\,.
  \end{equation}
  The tensors $\{ \mC_{11},\, \mC_{88} \}$, $\{ \mC_{11}',\, \mC_{88}' \}$ and $\{\mC_{\bar{3}3},\, \mC_{6\bar{6}} \}$
  are mutually orthogonal.

  Concerning flavor, a four-quark state made of $cq\bar{q}\bar{c}$ allows for
  several flavor wave functions. For further use we collect the combinations
  \begin{equation}\label{flavor}
  \begin{split}
     \mF_0 &=  cu\bar{u}\bar{c} + cd\bar{d}\bar{c}\,, \\
     \mF_1 &=  [cu]\{\bar{u}\bar{c}\} + [cd]\{\bar{d}\bar{c}\}\,, \\
     \mF_2 &=  \{cu\}[\bar{u}\bar{c}] + \{cd\}[\bar{d}\bar{c}]\,,
  \end{split}
  \end{equation}
  where $\{ \dots \}$ and $[ \dots ]$ denotes symmetrization and antisymmetrization,
  respectively.

  The Dirac part $\Gamma^\mu_\text{D}$ depends on the total momentum $P$ and the relative momenta $p$, $q$ and $k$. They
  are related to the individual (outgoing) quark momenta $p_i$ via
  \begin{equation} \renewcommand{\arraystretch}{2.5} \label{eq_partitioning_parameters_introduced}
  \begin{array}{rl}
  p_1 & \!\!= \displaystyle \frac{k+q-p}{2} + \sigma_1 P,\\
  p_2 & \!\!= \displaystyle \frac{k-q+p}{2} + \sigma_2 P,
  \end{array}\;\;
  \begin{array}{rl}
  p_3 &\!\!= \displaystyle \frac{-k+q+p}{2} + \sigma_3 P, \\
  p_4 &\!\!= \displaystyle \frac{-k-q-p}{2} + \sigma_4 P,
  \end{array}
  \end{equation}
  where $0\leq \sigma_i \leq 1$ is a set of momentum partitioning parameters
  satisfying $\sum_{i=1}^{4} \sigma_i=1$.
  The most general decomposition of $\Gamma^\mu_\text{D}$ with one vector index and four Dirac indices
  involves 768 linearly independent tensors $\tau^\mu_i$,
  which are collected in Appendix~\ref{sec:tensorbasis}:
  \begin{equation}\label{eq_amplitude}
  \Gamma^\mu_\text{D}(p,q,k,P) = \sum_{i=1}^{768} f_i(\Omega)\,\tau^\mu_i(p,q,k,P)\,.
  \end{equation}
  The scalar dressing functions $f_i$ depend on the nine Lorentz invariants obtained by combining
  the four-vectors $p$, $q$, $k$ and $P$: $\Omega= \{ p^2, q^2, k^2, p\cdot q, \dots \}$, with $P^2=-M^2$ fixed.
  Herein lies the main challenge in solving the tetraquark BSE numerically: Eq.~\eqref{eq_bse}
  produces a set of coupled integral equations for the dressing functions $f_i(\Omega)$
  which depend on nine independent variables each. This complexity makes calculations tedious and numerically challenging.

  In Ref.~\cite{Eichmann:2015cra} these nine Lorentz invariants were mapped
  onto a set of permutation-group variables~\cite{Eichmann2015}, which form a singlet $\mS_0$ and
  a doublet $\mD$,
  \begin{align}
  \mS_0 = \frac{p^2 + q^2 + k^2}{4}\,, \quad
  \mD= \frac{1}{4 \mS_0}
  \begin{pmatrix}
  \sqrt{3}(q^2-p^2) \\
  p^2+q^2-2 k^2
  \end{pmatrix},
  \label{eq_s4}
  \end{align}
  as well as two triplets $\mT_0$ and $\mT_1$, under transformations of the permutation group $S_4$.
  This ordering scheme allows one to take into account or discard the dependence of the amplitude
  on groups of variables (the members of the multiplets) without destroying
  its symmetries and turned out to be crucial with respect to numerical feasibility.

  In Ref.~\cite{Eichmann:2015cra} it was found that light scalar tetraquarks mainly depend on
  the three variables encoded in $\mS_0$ and $\mD$.
  Retaining $\mS_0$ only, their masses are in the ballpark of what one would naively expect for a state made of four quarks
  ($\sim 1500$ MeV for four light quarks). However,
  the four-quark BSE dynamically and self-consistently generates intermediate meson-meson and \mbox{$dq$-$\conjg{dq}$}
  poles in the dressing functions $f_i$ corresponding to the $(12)(34)$, $(13)(24)$ and $(14)(23)$ topologies.
  These poles appear in the Mandelstam plane formed by the doublet variables $\mD$ and produce decay thresholds.
  Due to the lightness of the pions and kaons as a consequence of spontaneous chiral symmetry breaking,
  they induce a sizeable mass shift by almost 1 GeV, such that the bound state around 1500 MeV turns into
  a resonance at a mass scale of about 400-500 MeV.
  Therefore, the implicit resonance mechanism and the possibility of a tetraquark to decay
  into light pseudoscalar mesons reduces its mass and leads to a mass pattern which
  is similar to the mass ordering for the $\sigma$, $\kappa$ and $a_0/f_0$ observed in experiment.

  While the permutation-group method greatly simplifies the problem, the dynamical creation of intermediate
  two-body poles still causes numerical complications.
  The eigenvalues of Eq.~\eqref{eq_bse} are calculated below all thresholds where the poles do not yet
  enter in the integration domain, but the rapid variation of the dressing functions in their vicinity makes numerical calculations cumbersome.
  Moreover, tracking the resonance locations in the complex plane would require to go above those thresholds
  using contour deformations~\cite{Williams:2018adr,Miramontes:2018omq,Eichmann:2019dts},
  where one additionally has to circumvent dynamically generated moving poles that only emerge in the solution of the equation itself.
  For these reasons we aim for another simplification in what follows and
  absorb the intermediate particle poles into the tensor basis of the amplitude.

\subsection{Physically motivated tensor basis}

 To describe an axialvector tetraquark with quark content $cq\bar{q}\bar{c}$ and quantum numbers $I(J^{PC})=0(1^{++})$
 in terms of meson-meson and \mbox{$dq$-$\conjg{dq}$} compositions, we consider the following combinations
 based on the two-particle decay modes of the X(3872) \cite{Tanabashi:2018oca}:
 \begin{itemize}
 \item $D^0_{ac} \conjg{D}^{*0}_{bd} + D^{*0}_{ac} \conjg{D}^0_{bd} + D^+_{ac} D^{*-}_{bd} + D^{*+}_{ac} D^{-}_{bd}$ \,,
 \item $J/\Psi_{ad} \,\omega_{bc}$\,,
 \item $S_{ab} A_{cd} + A_{ab} S_{cd}$\,.
 \end{itemize}
 The first is the composition in terms of a $D\conjg{D}^\ast$ molecule, which is motivated by the proximity of the $X(3872)$ to the $D\conjg{D}^\ast$ threshold.
 By contrast, the option $D\conjg{D}$ can produce $J=1$ only with non-vanishing orbital angular momentum.
 Second, a hadrocharmonium configuration containing a $J/\Psi$ with $0(1^{--})$
 would require a light pseudoscalar state with exotic quantum numbers $0(0^{--})$. Since such
 a state is not observed, this leaves the vector state $\omega$ with $0(1^{--})$ as a possible partner.
 Finally, viewed as a \mbox{$dq$-$\conjg{dq}$} state, the `good' diquarks are the scalar diquarks ($S$)
 with $J^P=0^+$ and the axialvector diquarks ($A$) with $J^P=1^+$, which are both in a color-antitriplet
 configuration so that the color-singlet tensor is $\mC_{\bar{3}3}$.

 Based on this, we approximate the Dirac-color-flavor structure of the BS amplitude as follows:
 \begin{equation}\label{amp-approx}
    \Gamma^\mu(p,q,k,P)  \approx \sum_{i=1}^3 f_i(\mS_0)\,T_i^\mu\,,
 \end{equation}
 where each $T_i^\mu$ corresponds to one of the compositions above:
 \begin{equation} \label{T1-T3}
 \begin{split}
    T_1^\mu &= R_1^\mu\,\mC_{11}\, \mF_0\,, \\[1mm]
    T_2^\mu &= R_2^\mu\, \mC_{11}'\, \mF_0\,, \\[1mm]
    T_3^\mu &= (R_3^\mu - R_4^\mu)\,\mC_{\bar{3}3}\, \mF_0\,.
 \end{split}
 \end{equation}
 Here, $\mC_{11}$, $\mC_{11}'$ and $\mC_{\bar{3}3}$ are the color tensors defined in Eqs.~(\ref{color-11}--\ref{color-33})
 and $\mF_0$ is the $I=0$ flavor tensor from Eq.~\eqref{flavor}.
 With the meson flavor wave functions given by
 $\{ D^0, \conjg{D}^0, \,D^+, \,D^-, \,J/\psi, \,\omega \}$ $\sim$
 $\{ c\bar{u}, u\bar{c}, c\bar{d}, d\bar{c}, c\bar{c}, u\bar{u} + d\bar{d} \}$,
 $\mF_0$ emerges automatically in the construction of the meson-meson components.
 The Dirac parts $R_i^\mu$ are
 \begin{align}
    R_1^\mu &= \left[ P(m_D,m_D^\ast)\,\gamma_5 \otimes \gamma^\mu_\perp - P(m_D^\ast,m_D)\,\gamma^\mu_\perp   \otimes \gamma_5\right]_{ac,bd}\,, \nonumber \\
    R_2^\mu &= \left[P(m_{J/\Psi},m_{\omega})\,\gamma^\alpha \otimes \gamma^\beta \hat{P}^\nu \epsilon^{\alpha \beta \nu \mu}\right]_{ad,bc} \,, \nonumber\\
    R_3^\mu &= \left[P(m_S,m_A)\,\gamma_5 C  \otimes C^T\gamma_\perp^\mu\right]_{ab,cd} \,, \label{R-Dirac} \\[1mm]
    R_4^\mu &= \left[P(m_A,m_S)\,\gamma_\perp^\mu C \otimes C^T \gamma_5 \right]_{ab,cd} \,, \nonumber
 \end{align}
 where $\gamma^\mu_\perp = \gamma_\mu- \hat{\slashed{P}}\,\hat{P}^\mu$ is the transverse $\gamma-$matrix,
 $\hat{P}$ the normalized total momentum,
 $C=\gamma_4\gamma_2$ the charge conjugation matrix and $T$ denotes a matrix transpose.
 The multi-indices $(a,b,c,d)$ stand for Dirac indices as well as momentum labels.

 In Eq.~\eqref{R-Dirac} we have absorbed the two-body poles, which would  emerge dynamically in the solution
 of the four-body equation, directly into the tensor structure by defining
 \begin{align*}
 P(m_1, m_2)_{ab,cd} &= \frac{1}{(p_a+p_b)^2 + m_1^2}
 \frac{1}{(p_c+p_d)^2 + m_2^2}\,.
 \end{align*}
 Given that this form captures all relevant momentum dependencies beyond the symmetric variable $\mS_0$,
 which turns out to be a good approximation for the light scalar tetraquarks~\cite{Eichmann:2015cra},
 the remaining dressing functions $f_i$  depend on $\mS_0$ only.
 The pole masses $m_D$, $m_{D^*}$, $m_{J/\psi}$, $m_\omega$, $m_S$ and $m_A$
 are calculated from the corresponding two-body BSEs in
 rainbow-ladder truncation~\cite{Eichmann:2016yit,Serna:2017nlr,Hilger:2017jti}
 and collected in Table~\ref{tab:rl}. The explicit form of the two-body interaction used in the
 BSE for the heavy-light mesons but also in the four-body BSE is given in Eq.~(3.96) of the review
 Ref.~\cite{Eichmann:2016yit}. We use the typical value $\Lambda=0.72$ GeV for the scale parameter,
 matched to reproduce the experimental value of the pion decay constant $f_\pi$, and $\eta=1.8 \pm 0.2$.
 We work in the isospin symmetric limit where $m_{D^+} = m_{D^-} = m_{D^0}$. The charm quark mass
 is determined by the condition that the sum $m_D+m_{D^*}$ equals the sum of the experimental masses
 \cite{Tanabashi:2018oca}. The strange quark mass is determined analogously for $m_{D_s}+m_{D_s^*}$.

 \begin{table}[t]\renewcommand{\arraystretch}{1.1}
 \begin{tabular}{ c @{\;\;} | @{\;\;} c @{\;\;} | @{\;\;} c @{\;\;} c @{\;\;} | @{\;\;}c @{\;\;} c }
             & $m_{\bar{q}}$        & $m_{PS}$ & $m_V$    & $m_S$    & $m_A$    \\ \hline\hline \rule{-0.0mm}{0.35cm}
  $q\bar{q}$ &     3.7              &  138(3)  &  732(1)  &  802(77) &  999(60) \\
  $c\bar{q}$ &     3.7              & 1802(2)  & 2068(16) & 2532(90) & 2572(8)  \\
  $c\bar{s}$ &      91              & 1911(3)  & 2169(14) & 2627(82) & 2666(7)  \\
  $c\bar{c}$ &     795              & 2792(6)  & 2980(6)  & 3382(15) & 3423(8)
 \end{tabular}
 \caption{Rainbow-ladder results for $q\bar{q}$, $c\bar{q}$, $c\bar{s}$ and $c\bar{c}$ meson and diquark masses (in MeV).
          $m_{\bar{q}}$ is the input current-quark mass at a renormalization point $\mu=19$ GeV in a MOM scheme.
          The column $m_{PS}$ contains  the masses of $\pi$, $D$, $D_s$ and $\eta_c$,
          the column $m_V$ those of $\rho/\omega$, $D^*$, $D_s^*$ and $J/\psi$,
          and the columns $m_S$ and $m_A$ list the corresponding diquark masses. The errors quoted
          are obtained by varying the parameter $\eta=1.8 \pm 0.2$.
          \label{tab:rl}}
 \end{table}

 In principle one could systematically proceed and construct a complete basis
 for the tetraquark amplitude with entangled Dirac, color and flavor tensors
 from all possible meson and diquark channels, and finally also restore the full momentum dependence of the dressing functions $f_i(\Omega)$.
 Our assumption here is that the amplitude is dominated
 by the three `physical' channels $D\conjg{D}^*$, $J/\psi\,\omega$ and \mbox{$dq$-$\conjg{dq}$}
 and that all momentum dependencies of the amplitude except those in $\mS_0$ can be
 absorbed in the $T_i^\mu$.
 This allows us to set $\mD = \mT_0 = \mT_1=0$ on the
 external momentum grid on the l.h.s. of the tetraquark BSE~\eqref{eq_bse}, which
 in the rest frame of the total momentum $P$ entails
\begin{equation}
   \left\{ k^\mu, \, p^\mu, \, q^\mu \right\} =  \frac{2 \mS_0}{\sqrt{3}}\left\{ e_1^\mu, \, e_2^\mu, \, e_3^\mu \right\}, \quad
   P^\mu = iM\,e_4^\mu\,,
\end{equation}
 where $e_i^\mu$ are the Euclidean unit vectors.
 The r.h.s of~\eqref{eq_bse} still samples the full domain of $\Omega$ under the integral.

\begin{figure*}[t]
\includegraphics[width=1\columnwidth]{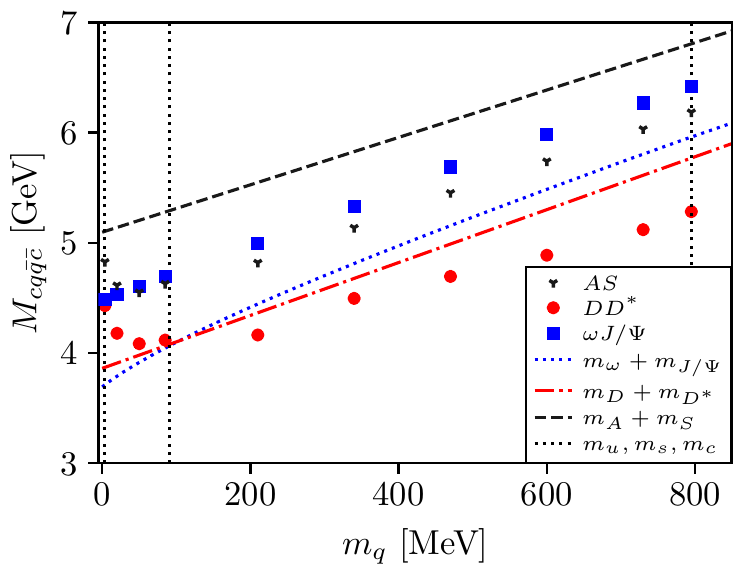}
\includegraphics[width=1\columnwidth]{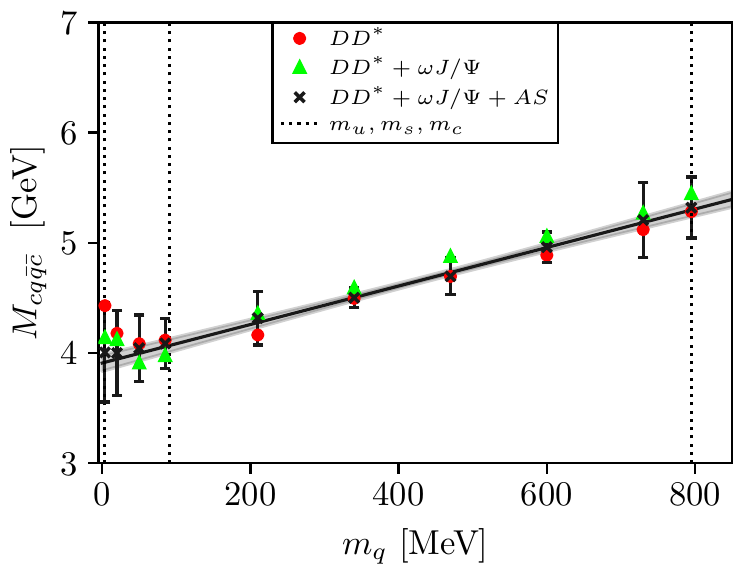}
\caption{Mass of the $I(J^{PC})=0(1^{++})$ four-quark state as a function of the current-quark mass. \textit{Left:} Solutions for the individual
         $D\conjg{D}^*$, $J/\psi\,\omega$ and diquark-antidiquark components together with their respective thresholds.
         \textit{Right:} results obtained from including one ($D\conjg{D}^*$), two and all three channels. The error bars combine the extrapolation error
         with the error obtained by varying the momentum partitioning parameter $\zeta$.
         The black line is a fit to the data points together with an error band.
          }  \label{fig:results}
\end{figure*}

Concerning the \mbox{$dq$-$\conjg{dq}$} flavor structure in Eq.~(\ref{T1-T3}), one can alternatively consider the tensor
 \begin{equation}\label{diquark-T3}
   \widetilde T_3^\mu = (  R_4^\mu \,\mF_2
    -
    R_3^\mu \,\mF_1 )\,\mC_{\bar{3}3}\,,
 \end{equation}
where the Pauli antisymmetry of the diquarks (which would hold for exact $SU(4)$ flavor symmetry)
 is explicit. Combined with the antisymmetric color tensor $\sim \varepsilon_{ABC}$,
 the antisymmetric tensor $\gamma_5 C$ for a scalar diquark is matched with an antisymmetric
 flavor structure and the symmetric tensor $\gamma^\mu C$ for the axialvector diquarks with a symmetric
 flavor wave function. In contrast to Eq.~\eqref{T1-T3}, however, after plugging Eq.~\eqref{diquark-T3} into the BSE and taking
 flavor traces the diquarks decouple from the meson-meson configurations in this case,
 so that both can be solved independently and yield two different states.
 This setup requires no separate discussion since
 the decoupled \mbox{$dq$-$\conjg{dq}$} state is identical for $T_3$ and $\widetilde{T}_3$ and
 the equations for the meson-meson configuration are the same when
 dropping the diquarks in the construction of the amplitude in Eq.~\eqref{amp-approx},
 which we will do below for comparison.

\section{Results}\label{secIII}

  Before discussing the results, we briefly summarize our setup.
  We solve the four-body equation depicted in Fig.~\ref{fig_q4_bse}
  with the rainbow-ladder kernel including all permuations, where
  the dressed light and charm-quark propagators are obtained from their DSEs.
  We approximate the structure of the tetraquark amplitude by its three dominant
  components in Eq.~\eqref{T1-T3}, which has
  the advantage of reducing the complexity of the four-body equation while
  the system still dynamically decides which of the three configurations
  -- heavy-light meson, hadrocharmonium or \mbox{$dq$-$\conjg{dq}$} -- is most important.

  In Fig.~\ref{fig:results} we track the tetraquark mass as a function of the light current-quark mass $m_q$,
  which we vary from the charm-quark mass ($cc\bar{c}\bar{c}$) down to the up/down quark mass ($cq\bar{q}\bar{c}$).
  At each quark mass and for each setup we calculate the eigenvalues $\lambda_i(P^2)$ of the kernel of the tetraquark BSE~\eqref{eq_bse}
  below the lowest-lying two-particle decay threshold, i.e., for $P^2 > - M_\text{thr}^2$ along the real axis.
  We track the largest eigenvalue as a function of $P^2$ and read off the mass $M$ of the ground state from the condition $\lambda_0(P^2=-M^2) = 1$.
  In cases where this condition is not satisfied below the threshold, we extrapolate the eigenvalue to obtain an estimate for the real part of the resonance mass,
  although one should be cautious when interpreting results obtained from extrapolating over thresholds~\cite{Hanhart2014}.
  As explained earlier, we cannot yet access the whole complex $P^2$ plane due to restrictions imposed by the intermediate meson and diquark poles
  as well as the poles in the complex plane of the quark propagator, which would require contour deformations.

  In the left panel of Fig.~\ref{fig:results} we plot the results obtained for each of the three configurations in Eq.~\eqref{amp-approx} separately:
  $D\conjg{D}^*$, $J/\psi\, \omega$ and \mbox{$dq$-$\conjg{dq}$}, together with their respective thresholds.
  The thresholds for $D\conjg{D}^*$ and $J/\psi\, \omega$ are relatively close to each other, whereas the sum
  of the scalar and axialvector diquark masses obtained from their BSEs is larger by almost 1 GeV.
  This already provides a first indication that the diquark contributions may be subleading compared to the meson-meson components
  simply due to their larger masses, similarly as in the case of the light scalar mesons~\cite{Eichmann:2015cra}.

  This is indeed what we observe in the left panel of Fig.~\ref{fig:results}: Solved for each individual tensor $T_1$, $T_2$ or $T_3$ alone,
  it turns out that the $D\conjg{D}^*$ component has the lowest mass, followed by the \mbox{$dq$-$\conjg{dq}$} and the $J/\psi \,\omega$ components.
  The \mbox{$dq$-$\conjg{dq}$} mass is always below its threshold, whereas the $D\conjg{D}^*$ mass lies below threshold only
  above the strange-quark mass and the $J/\psi\,\omega$ mass comes out above its threshold. In the right panel of Fig.~\ref{fig:results} we show again the result for the $D\conjg{D}^*$
  component $T_1$ together with the results including both $T_1$ and $T_2$ and the full result
  with all three tensors. Indeed one can see that the addition of the \mbox{$dq$-$\conjg{dq}$}
  tensors and the $J/\psi\,\omega$ component has little effect and the mass of the state is
  essentially determined by the $D\conjg{D}^*$ component alone.

  To provide an estimate of the error of the calculation we vary the momentum partitioning parameters
  in Eq.~\eqref{eq_partitioning_parameters_introduced}. Had we included all kinematic variables,
  i.e. all permutation group multiplets discussed above, the results would be independent of the
  momentum partitioning; however, with a smaller subset this is no longer the case. In practice we
  choose $\sigma_1 = \sigma_4 = \tfrac{1}{2} - \zeta$ and $\sigma_2 = \sigma_3 = \zeta$,
  which leaves one parameter $\zeta$ that quantifies the fraction of the momentum $P$
  assigned to the $c\bar{c}$ and $q\bar{q}$ pairs. We then optimize that value to minimize
  the distance to the thresholds and vary $\zeta$ in the vicinity of its optimal value.
  The errors obtained in this case are shown in the first row of Table~\ref{tab_error_band}.
  One clearly sees that the error increases with decreasing light quark mass $m_q$.
  The second row shows the error obtained by optimizing the momentum partitioning
  without restricting the $\sigma_i$, which has a similar but even larger effect.
  The rise of the tetraquark masses in Fig.~\ref{fig:results} towards smaller quark masses
  (even for points below thresholds) is thus an artefact of the reduced kinematics,
  which induces an error of at least $10 \%$ at the physical $u/d$ mass.

  Within errors, all three results of Fig.~\ref{fig:results} are the same and dominated by the
  heavy-light meson $D\conjg{D}^*$ component, whereas the \mbox{$dq$-$\conjg{dq}$} and
  $J/\psi\,\omega$ components are almost negligible.
  The black solid line in the figure is obtained by fitting the linear expression
  \begin{equation}
     M = c_0 + c_1\,m_q
  \end{equation}
  to the six data points at the larger current-quark masses where the tetraquark is definitely a bound state.
  The grey band indicates the combined errors from the momentum partitioning and --
  for the four lowest quark masses -- the extrapolation error.
  At the physical $u/d$ quark mass, the fit yields
  \begin{equation}
     M_{1^{++}}^{cq\bar{q}\bar{c}} =
3916(74)
    \,\text{MeV}
  \end{equation}
  for the mass of the axialvector tetraquark in good agreement with the mass of the X(3872).
  In addition we find
  \begin{equation}
     M_{1^{++}}^{cs\bar{s}\bar{c}} =
4068(61)
     \,\text{MeV}
  \end{equation}
  for the mass of a putative four-quark state with charm and strange quarks. Note that from
  Fig.~\ref{fig:results} one cannot read off the mass of an all-charm tetraquark since the
  implemented symmetries among the various constituents are no longer appropriate when all
  four quarks are equal.

\begin{table}\renewcommand{\arraystretch}{1.2}
\begin{tabular}{c|r@{\;\;}r@{\;\;}r@{\;\;}r@{\;\;}r@{\;\;}r@{\;\;}r@{\;\;}r@{\;\;}r@{\;\;}c}
$m_q$\,[MeV] & 3.7 & 20	& 50 &	85 & 210 & 340 & 470 & 600 & 730 & 795 \\ \hline\hline \rule{-0.0mm}{0.35cm}
$\Delta M/ M$ $[\% ]$ & 7.4& 4.7& 4.2& 2.9& 1.7& 1.0& 0.5& 0.3& 0.1& 0.02
 \\
$\Delta M/ M$ $[\% ]$&
10.1 & 9.2 & 7.4 & 5.5 & 5.8 & 2.0 & 3.5 & 2.8 & 6.6 & 5.3
\end{tabular}
\caption{Error estimates from the momentum partitioning for one parameter $\zeta$ (row 1)
         and without restricting the $\sigma_i$ (row 2). \label{tab_error_band}}
\end{table}

\section{Conclusions}\label{sec:conclusion}

When solving the four-body equation for a system of heavy-light quark flavours in
the $I(J^{PC})=0(1^{++})$ channel, we find a similar behaviour as for the light scalar
meson octet discussed in Ref.~\cite{Heupel:2012ua,Eichmann:2015cra}: The ground state
is dominated by a strong meson-meson component. For the light scalar mesons this component
has pseudoscalar quantum numbers and is strongly affected by the effects of dynamical
chiral symmetry breaking. For the case considered here, it is a combination of heavy-light
pseudoscalar and vector components, namely the $D\conjg{D}^*$ combination. In both cases,
diquark-antidiquark components in the wave function are negligible. For the case of heavy-light
four-quark states one can in addition identify a hadro-charmonium contribution to the
wave function which, however, is also sub-dominant as compared to the dominating
heavy-light $D\conjg{D}^*$ component.

The precision of our calculation is not good enough
to decide whether the resulting axialvector four-body state is a bound meson molecule or not --- to
this end we would need to be able to determine its mass on the several-MeV level, which is
not possible with the tools at hand. Nevertheless, within error bars, we find a state in the
correct mass range to be identified with the X(3872). In order to corroborate our findings we
need to gain precision using appropriate techniques to deal with the analytic structure of the
four-body equation beyond extrapolations. Furthermore it would be very interesting to include
a $c\bar{c}$ component in the wave function in order to address mixing with ordinary charmonia.
These improvements are subject to future work.

\smallskip
{\bf Acknowledgements}\\
We are grateful to Christoph Hanhart, Soeren Lange and Marc Wagner for discussions.
This work was supported by the Helmholtz International Center for FAIR within
the LOEWE program of the State of Hesse, by the DFG grant FI 970/11-1
and by the FCT Investigator Grant IF/00898/2015.

\begin{appendix}

\section{Tensor basis}\label{sec:tensorbasis}

   The axialvector tetraquark amplitude in Eq.~\eqref{eq_amplitude} depends on 768 Dirac tensor basis elements.
   To derive them, we first recapitulate the construction of the basis for a scalar tetraquark~\cite{Eichmann:2015cra}.
   By orthogonalizing the momenta $p,q,k,P$ one obtains
   four orthonormal momenta $n_i^\mu$ ($i=1\dots 4$) that are mutually transverse. The set
   \begin{equation}\label{basis-set-1}
       \{ \mathds{1}\,, \;\; \slashed{n}_i\,, \;\; \slashed{n}_i \,\slashed{n}_j\,, \;\; \slashed{n}_i\,\slashed{n}_j\,\slashed{n}_k\,, \;\; \slashed{n}_i\,\slashed{n}_j\,\slashed{n}_k\,\slashed{n}_l \}
   \end{equation}
   with $i<j<k<l$ consists of 16 elements;
   commutators are not necessary because $\slashed{n}_i\,\slashed{n}_j = -\slashed{n}_j\,\slashed{n}_i$.
   Taking all tensor products of Eq.~\eqref{basis-set-1} with itself yields $256$ linearly independent tensor structures.
   No further $\gamma-$matrices are necessary because they can be reconstructed from the unit vectors:
   \begin{equation}\label{delta-gamma}
      \delta^{\mu\nu} = \sum_{i=1}^4 n_i^\mu\,n_i^\nu\,, \qquad
      \gamma^\mu = \sum_{i=1}^4 n_i^\mu\,\slashed{n}_i\,.
   \end{equation}

   To further simplify~\eqref{basis-set-1} we define the pseudoscalar
   \begin{equation}
       \epsilon = \varepsilon^{\mu\nu\rho\sigma}\,n_1^\mu\,n_2^\nu\,n_3^\rho\,n_4^\sigma
   \end{equation}
   which can take values $\epsilon = \pm 1$.
   From the relation
   \begin{equation}
       \frac{1}{24}\,[\gamma^\mu, \gamma^\nu, \gamma^\alpha, \gamma^\beta] = -\gamma_5\,\varepsilon^{\mu\nu\alpha\beta}\,,
   \end{equation}
   where the four-commutator is the fully antisymmetrized product of four $\gamma-$matrices,
   one obtains
   \begin{equation}
       \slashed{n}_1\,\slashed{n}_2\,\slashed{n}_3\,\slashed{n}_4 = -\epsilon \gamma_5 \,.
   \end{equation}
   In this way all elements in~\eqref{basis-set-1} with three or four slashes can be reduced to those with two at most,
   so we can write it as $\{ \mathds{1}\,, \, \slashed{n}_i\,, \, \slashed{n}_4\,, \,\slashed{n}_i\,\slashed{n}_4 \} \, \times \, \Omega_\omega$
   with $\Omega_1=\mathds{1}$, $\Omega_2=\epsilon\gamma_5$ and $i=1,2,3$.
   If $n_4^\mu = \hat{P}^\mu$ denotes the normalized total momentum and
   we express $\slashed{n}_4$ in terms of the positive/negative-energy projectors $\Lambda_\pm =(\mathds{1} \pm \slashed{n}_4)/2$,
   then~\eqref{basis-set-1} becomes $\{ \mathds{1}\,, \, \slashed{n}_i\} \, \times \,\Lambda_\lambda \, \times \,\Omega_\omega$.

   A complete, orthonormal, covariant, 256-dimensional positive-parity basis for the BS amplitude
   of a scalar tetraquark is then given by
   \begin{equation}\label{basis-full-1}
       \tau_n(p,q,k,P) = \Gamma_j\,\Lambda_\lambda \,\Omega_\omega \,\gamma_5 C\otimes C^T\gamma_5\,\Omega_{\omega'}\,\Lambda_{\lambda'}\,\Gamma_k
   \end{equation}
   with $\Gamma_j \in \{ \mathds{1}, \, \slashed{n}_1, \slashed{n}_2, \slashed{n}_3 \}$.
   We inserted the combination $\gamma_5 C \otimes C^T\gamma_5$ for the \mbox{$dq$-$\conjg{dq}$} topology (12)(34);
   all further structures such as $\gamma^\mu C\otimes C^T\gamma^\mu$ but also those in the meson-meson topologies are linearly dependent.

   Because the $\gamma-$matrices can be reduced to the unit vectors and their slashes according to Eq.~\eqref{delta-gamma},
   an orthonormal basis for the axialvector tetraquark simply follow from
   attaching $\epsilon n_i^\mu$ with $i=1,2,3$ to~\eqref{basis-full-1}.
   $n_4^\mu$ cannot appear because the tetraquark must be transverse
   in the total momentum. This yields $3 \times 256 = 768$ linearly independent and covariant Lorentz-Dirac tensors.

             \begin{table}[t]

             \begin{equation*} \renewcommand{\arraystretch}{1.2}
             \begin{array}{ l @{\quad} |  @{\quad} c  @{\quad} c @{\quad} c @{\quad} c @{\;\;}  }

                                                & s  &  p  &  d  &  f   \\[1mm] \hline\hline \rule{-0.0mm}{0.4cm}

                 n_i^\mu                        & 0  &  3  &  0  &  0   \\
                 n_i^\mu\,n_j^\alpha            & 1  &  3  &  5  &  0   \\
                 n_i^\mu\,n_k^\beta             & 1  &  3  &  5  &  0   \\
                 n_i^\mu\,n_j^\alpha\,n_k^\beta & 1  &  9  &  10  &  7

             \end{array}
             \end{equation*}

               \caption{Eigenfunctions of the orbital angular momentum operator
                        obtained from combinations of the three unit vectors $n_i^\mu$,
                        which correspond to the relative momenta $p$, $q$ and $k$.}
               \label{spdf}

             \end{table}

       In practice we are interested in partial-wave bases whose tensors are eigenstates
       of the total quark spin and orbital angular momentum in the tetraquark's rest frame.
       The construction is analogous to the nucleon's Faddeev amplitude~\cite{Eichmann:2011vu}.
       The eigenvalues of the quark spin can take values $s=0,1,2$, which can combine with $l=0,1,2,3$ to produce
       total angular momentum $J=1$.
       The orbital angular momentum operator $L^2$ only acts on the unit vectors $n_i^\mu$ with $i=1,2,3$ but not on the
       total momentum $\sim n_4^\mu$, and it leaves Lorentz scalars as well as $\epsilon$ invariant.
       Thus we only need to consider the combinations $n_i^\mu\,\{ \mathds{1}\,, \, \slashed{n}_j \} \otimes \{ \mathds{1}\,, \, \slashed{n}_k \}$.
       $L^2$ does also not act on the Dirac structure, so the problem reduces to arranging
       \begin{equation}
          n_i^\mu\,, \quad
          n_i^\mu\,n_j^\alpha\,, \quad
          n_i^\mu\,n_k^\beta\,, \quad
          n_i^\mu\,n_j^\alpha\,n_k^\beta
       \end{equation}
       into combinations that are eigenfunctions of $L^2$ with eigenvalues $l(l+1)$.

       The resulting number of $s$, $p$, $d$ and $f$ waves is given in Table~\ref{spdf}.
       For example, the three unit vectors $n_i^\mu$ are $p$ waves because $L^2\,n_i^\mu = 2n_i^\mu$.
       The nine combinations $n_i^\alpha\,n_j^\beta$ produce one $s$ wave with $l=0$:
       \begin{equation}
           T^{\alpha\beta} = \sum_{i=1}^3 n_i^\alpha\,n_i^\beta = \delta^{\alpha\beta} - n_4^\alpha\,n_4^\beta\,,
       \end{equation}
       three $p$ waves with $l=1$:
       \begin{equation}
           n_i^\alpha\,n_j^\beta - n_j^\alpha\,n_i^\beta = \epsilon\,\varepsilon^{\alpha\beta\gamma\delta}\,n_k^\gamma\,n_4^\delta\,,
       \end{equation}
       where $\{ i,j,k\}$ is an even permutation of $\{1,2,3\}$, and five $d$ waves with $l=2$:
       \begin{equation*}
           n_i^\alpha\,n_i^\beta - \tfrac{1}{3}\,T^{\alpha\beta} \quad (i=2,3)\,, \quad
           n_i^\alpha\,n_j^\beta + n_j^\alpha\,n_i^\beta \quad (i\neq j).
       \end{equation*}
       Similarly, the 27 combinations $n_i^\alpha\,n_j^\beta\,n_k^\gamma$ produce one $s$ wave
       $\epsilon\,\varepsilon^{\alpha\beta\gamma\delta}\,n_4^\delta$ as well as further $p$, $d$ and $f$ waves.

       Putting these eigenfunctions back into the tensor basis yields the partial-wave decomposition
       of the amplitude, for example for the $s$ waves:
       \begin{alignat}{3}
          n_i^\mu\,\slashed{n}_j &\otimes \mathds{1}  &&\longrightarrow \;  T^{\mu\alpha}\,\gamma^\alpha \otimes \mathds{1} = \gamma^\mu_\perp \otimes \mathds{1}\,, \nonumber \\
          n_i^\mu\,\mathds{1} &\otimes \slashed{n}_k  &&\longrightarrow \;  T^{\mu\beta}\,\mathds{1}\otimes \gamma^\beta = \mathds{1} \otimes \gamma^\mu_\perp\,, \\
          n_i^\mu\,\slashed{n}_j &\otimes \slashed{n}_k &&\longrightarrow \; \epsilon\,\varepsilon^{\mu\alpha\beta\gamma}\,n_4^\gamma\,\gamma^\alpha\otimes\gamma^\beta\,. \nonumber
       \end{alignat}
       Combined with the 16 tensors $\Lambda_\lambda \,\Omega_\omega \,\gamma_5 C\otimes C^T\gamma_5\,\Omega_{\omega'}\,\Lambda_{\lambda'}$
       from Eq.~\eqref{basis-full-1}, this yields $48$ $s$-wave basis elements which are Fierz complete.
       The tensors in Eq.~\eqref{R-Dirac} project onto a subset of these.
       Since they carry $l=0$, their dressing functions capture the dominant momentum dependence of the
       axialvector tetraquark amplitude.

\end{appendix}

\bibliographystyle{apsrev4-1}
\bibliography{X_tetraquark2}

\end{document}